%% file: main.tex
\documentclass[twocolumn]{article}

\usepackage{framed,multirow}

\usepackage{amssymb}
\usepackage{latexsym}

\usepackage{url}
\usepackage{xcolor}

\usepackage{hyperref}

\usepackage{placeins}
\usepackage{graphicx}
\usepackage{subcaption}

\definecolor{newcolor}{rgb}{.8,.349,.1}

\usepackage{geometry} 
\title{The STOIC2021 COVID-19 AI challenge: applying reusable training methodologies to private data}

\input{authors.tex}

\begin{document}

\maketitle


\begin{abstract}
\input{sections/abstract}
\end{abstract}



\input{sections/intro}

\input{sections/method}
\input{sections/results}
\FloatBarrier
\input{sections/discussion}

\section*{Acknowledgments}
The European Regional Development Fund had no role in the study design, data collection, data analysis, data interpretation, or writing of the manuscript. Amazon Web Services funded algorithm evaluation, algorithm training for the Final phase, and prizes to the best performing teams. This study was endorsed by The Medical Image Computing and Computer Assisted Intervention (MICCAI) Society. The STOIC study \cite{revel2021study} was sponsored by Assistance Publique Hôpitaux de Paris and was funded by Fondation APHP pour la Recherche, Guerbet, Innothera, Fondation CentraleSupélec. For the STOIC study, General Electric Healthcare provided a 3D image visualization web application and Orange Healthcare a data repository.

\section*{Role of the funding resource}
The European Regional Development Fund had no role in the study design, data collection, data analysis, data interpretation, or writing of the manuscript. Amazon Web Services funded algorithm evaluation, algorithm training for the Final phase, and prizes to the best performing teams. 

\section*{Data statement}
Any collaborative research project led by an academic partner who requires access to the STOIC data shall be analyzed, validated, and authorized by the Steering Committee of STOIC. To this end, the academic partner shall send a document describing the research project to the Stoic Steering Committee at the following email address: \url{marie-pierre.revel@aphp.fr}, with the following subject: STOIC DATA ACCESS PERMISSION. After acceptance by the Steering Committee, the academic partner shall sign a specific agreement (Data Transfer Agreement - DTA) with AP-HP, who is legally responsible for the STOIC data as Sponsor of the STOIC research. Refer to the STOIC princeps paper \cite{revel2021study} for more information.

\section*{Declaration of generative AI in scientific writing}
During the preparation of this work the author(s) used ChatGPT in order to improve readability and language. After using this tool/service, the author(s) reviewed and edited the content as needed and take(s) full responsibility for the content of the publication.

\FloatBarrier
\bibliographystyle{plain}
\bibliography{bib/fullstrings,bib/diagnoweb,bib/diag,bib/newrefs}
\end{document}

%% file: authors.tex
\usepackage{authblk}

\author[1]{Luuk H. Boulogne \footnote{corresponding author: luuk.boulogne@radboudumc.nl}}
\author[2]{Julian Lorenz}
\author[2]{Daniel Kienzle}
\author[2]{Robin Schön}
\author[2]{Katja Ludwig}
\author[2]{Rainer Lienhart}

\author[3]{Simon Jégou}

\author[4]{Guang Li}
\author[4]{Cong Chen}
\author[4]{Qi Wang}
\author[4]{Derik Shi}

\author[5]{Mayug Maniparambil}

\author[6,2]{Dominik Müller}
\author[6]{Silvan Mertes}
\author[6]{Niklas Schröter}
\author[6]{Fabio Hellmann}
\author[6]{Miriam Elia}

\author[7,14]{Ine Dirks}
\author[7,14]{Matías Nicolás Bossa}
\author[7,14]{Abel Díaz Berenguer}
\author[7,14]{Tanmoy Mukherjee}
\author[7,14]{Jef Vandemeulebroucke}
\author[7,14]{Hichem Sahli}
\author[7,14]{Nikos Deligiannis}
\author[7,14]{Panagiotis Gonidakis}

\author[8]{Ngoc Dung Huynh}
\author[9]{Imran Razzak}
\author[8]{Reda Bouadjenek}

\author[10]{Mario Verdicchio}
\author[10]{Pasquale Borrelli}
\author[10]{Marco Aiello}

\author[1]{James A. Meakin}
\author[11]{Alexander Lemm}
\author[11]{Christoph Russ}
\author[11]{Razvan Ionasec}
\author[12,4]{Nikos Paragios}
\author[1]{Bram van Ginneken}
\author[13]{Marie-Pierre Revel Dubois}

\affil[1]{Radboud university medical center; P.O.Box 9101, 6500HB Nijmegen, The Netherlands}
\affil[2]{University of Augsburg, Universitätsstraße 2, 86159 Augsburg, Germany}
\affil[3]{Independent researcher}
\affil[4]{Keya medical technology co. ltd, Floor 20, Building A, 1 Ronghua South Road, Yizhuang Economic Development Zone, Daxing District, Beijing, PRC}
\affil[5]{ML-Labs, Dublin City University, N210, Marconi building, Dublin City University, Glasnevin, Dublin 9, Ireland}
\affil[6]{Faculty of Applied Computer Science, University of Augsburg}
\affil[7]{Vrije Universiteit Brussel, Department of Electronics and Informatics, Pleinlaan 2, 1050 Brussels, Belgium}
\affil[8]{Deakin University, Geelong, Australia}
\affil[9]{University of New South Wales, Sydney, Australia}
\affil[10]{IRCCS SYNLAB SDN, Naples, Italy}
\affil[11]{Amazon Web Services, Marcel-Breuer-Str. 12, 80807 München, Germany}
\affil[12]{TheraPanacea, 75004, Paris, France}
\affil[13]{Department of Radiology, Université de Paris, APHP, Hôpital Cochin, 27 rue du Fg Saint Jacques, 75014 Paris, France}
\affil[14]{imec, Kapeldreef 75, 3001 Leuven, Belgium}

%% file: sections/abstract.tex
Challenges drive the state-of-the-art of automated medical image analysis. The quantity of public training data that they provide can limit the performance of their solutions. Public access to the training methodology for these solutions remains absent. This study implements the Type Three (T3) challenge format, which allows for training solutions on private data and guarantees reusable training methodologies. With T3, challenge organizers train a codebase provided by the participants on sequestered training data. T3 was implemented in the STOIC2021 challenge, with the goal of predicting from a computed tomography (CT) scan whether subjects had a severe COVID-19 infection, defined as intubation or death within one month. STOIC2021 consisted of a Qualification phase, where participants developed challenge solutions using 2\,000 publicly available CT scans, and a Final phase, where participants submitted their training methodologies with which solutions were trained on CT scans of 9\,724 subjects. 
The organizers successfully trained six of the eight Final phase submissions. The submitted codebases for training and running inference were released publicly. The winning solution obtained an area under the receiver operating characteristic curve for discerning between severe and non-severe COVID-19 of 0.815. The Final phase solutions of all finalists improved upon their Qualification phase solutions. 

%% file: sections/intro.tex
\begin{figure*}[ht]
  \centering
  \includegraphics[width=.75\textwidth]{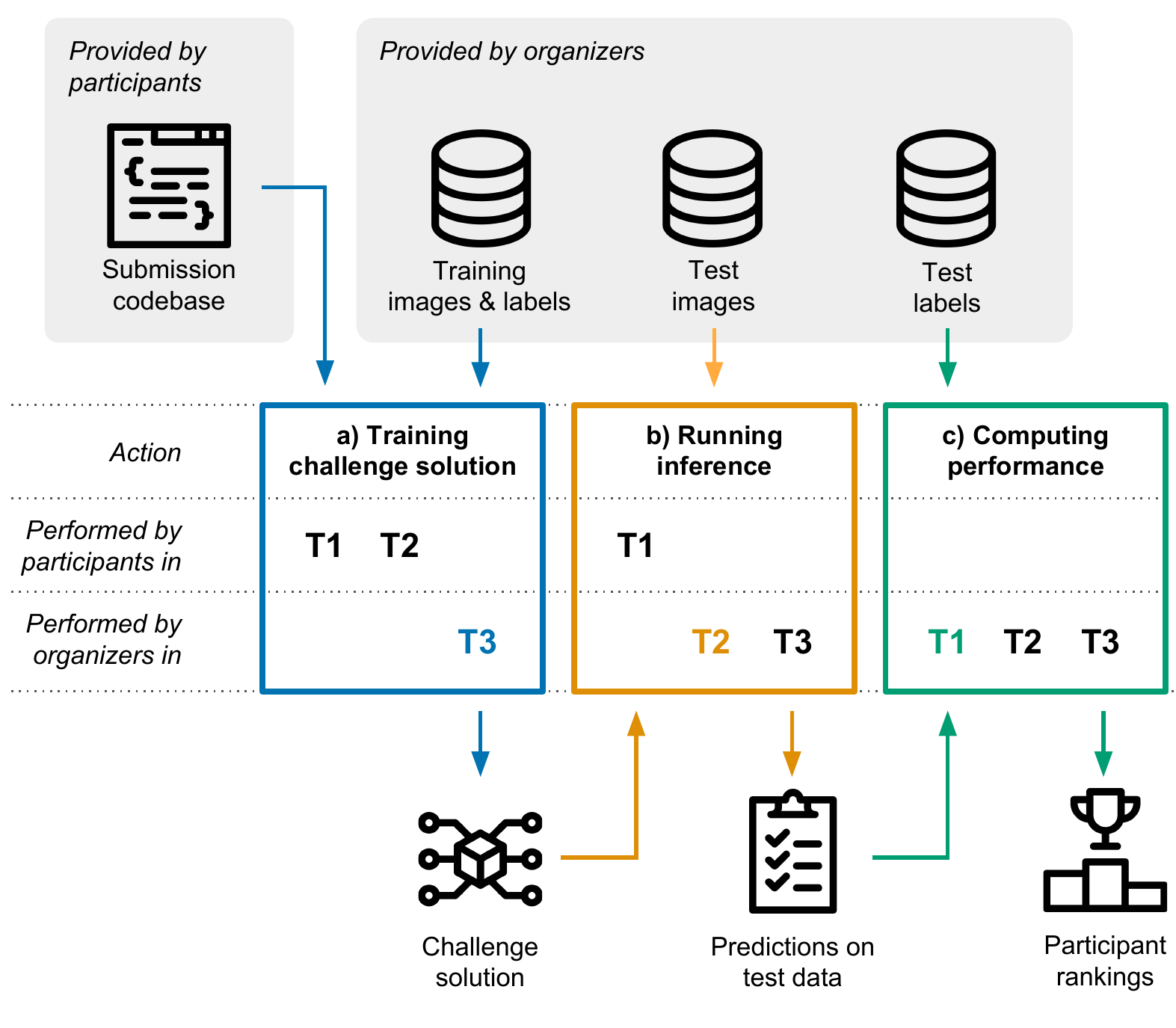}
  \caption{Schematic representation of the submission pipeline of challenges of Type One (T2), Type Two (T2), and Type Three (T3). a) A challenge solution is trained by applying a participants’ codebase to images and labels provided by the challenge organizers. With T1 and T2, participants perform this step. With T3, the challenge organizers perform training. b) The solution is applied to test images, producing predictions. The introduction of the T2 format allowed challenge organizers to perform this step. c) The resulting predictions are compared with test labels to compute the submission’s performance. Participants are ranked based on their performance. In all challenge types, the performance is computed by the organizers.}
  \label{fig:1}
\end{figure*}

\section{Introduction}
Grand challenges for medical image analysis aim to provide the best solutions to clinical problems that the field of artificial intelligence has to offer. The performance of these solutions can be limited by the quantity of data for model development that the challenge organizers release publicly. Although some recent challenges ensured that the winning solutions were readily available after the challenge had completed, \cite{Bult22,Aubr22,schirmer2021neuropsychiatric,da2022digestpath,ouyang2019analysis} reusability of the methods with which these solutions were trained was not enforced.

This work implements a challenge format that allows for training submissions on private data. This ensures that the winning solutions can easily be retrained on new datasets after the challenge has concluded.

We aim to demonstrate the effectiveness of this challenge format in the STOIC2021 challenge, available at \url{https://stoic2021.grand-challenge.org}. The focus of STOIC2021 was to produce fully automatic methods for discriminating between severe and non-severe COVID-19 subjects, with severe COVID-19 defined as death or intubation after one month. The challenge was organized with data from the STOIC project, \cite{revel2021study} a multi-center dataset that comprises CT scans of 10\,735 subjects. The STOIC project protocol can be accessed via ClinicalTrials.gov with identifier NCT04355507.

Through STOIC2021, this study provides the public release of CT scans of 2\,000 subjects suspected for COVID-19, along with RT-PCR results, disease severity at one month follow-up, age, and sex labels under a CC-BY-NC 4.0 licence.

The submission pipeline of a challenge generally consists of training a challenge solution, running inference with it on a test set, and using the resulting predictions to compute the submission’s performance. In this work, we define different challenge types by considering which steps are performed by challenge participants, and which steps are performed by challenge organizers. Figure \ref{fig:1} describes the challenge submission pipeline, previously used challenge formats that are referred to in this work as Type One (T1) and Type Two (T2), as well as the Type Three (T3) challenge format.

In T1 challenges, \cite{ouyang2019analysis,Anto21,Ehte17,lassau2020three,choi2022challenge,halabi2019rsna,ali2021deep,knoll2020advancing,porwal2020idrid,kim2021paip,fang2022adam,sun2021multi,sathianathen2022automatic,combalia2022validation,kavur2021chaos,hakim2021predicting,Hell19a,Bogu19,orlando2020refuge,Yang18a,hirvasniemi2023knee,arganda2015crowdsourcing,ivantsits2022detection,caicedo2019nucleus,simoes2020bciaut,Veta18,winzeck2018isles,marinescu2019tadpole,balagurunathan2021lung,de2021generalizability,bratholm2021community,bron2015standardized,Seti17,pan2019improving,cash2015assessing,kim2020challenge,qsm2021qsm,fu2020age,babier2021openkbp} participants perform inference on a publicly released test set themselves, which does not preclude them from meddling with their predictions. T2 challenges \cite{Bult22,Aubr22,schirmer2021neuropsychiatric,da2022digestpath,sun2022crowdsourcing,hatt2018first} solve this issue by requiring participants to submit functional algorithms. These can be made easily accessible to third parties \cite{Bult22,Aubr22,schirmer2021neuropsychiatric,da2022digestpath}, and generate reproducible results \cite{Bult22,sun2022crowdsourcing}.

We implement the Type Three (T3) challenge structure, which has only seen limited use in medical image analysis research \cite{schaffter2020evaluation}.  With T3, participants do not submit an algorithm for inference, but they instead submit a codebase for training and inference. The challenge organizers apply the codebase to the training set, generating the corresponding challenge solution. This allows for training on a combination of public and sensitive private training data. It guarantees that not only inference methods, but also training methods work out-of-the-box for third parties.

%% file: sections/method.tex
\begin{figure*}[!htb]
  \centering
  \includegraphics[width=.90\textwidth]{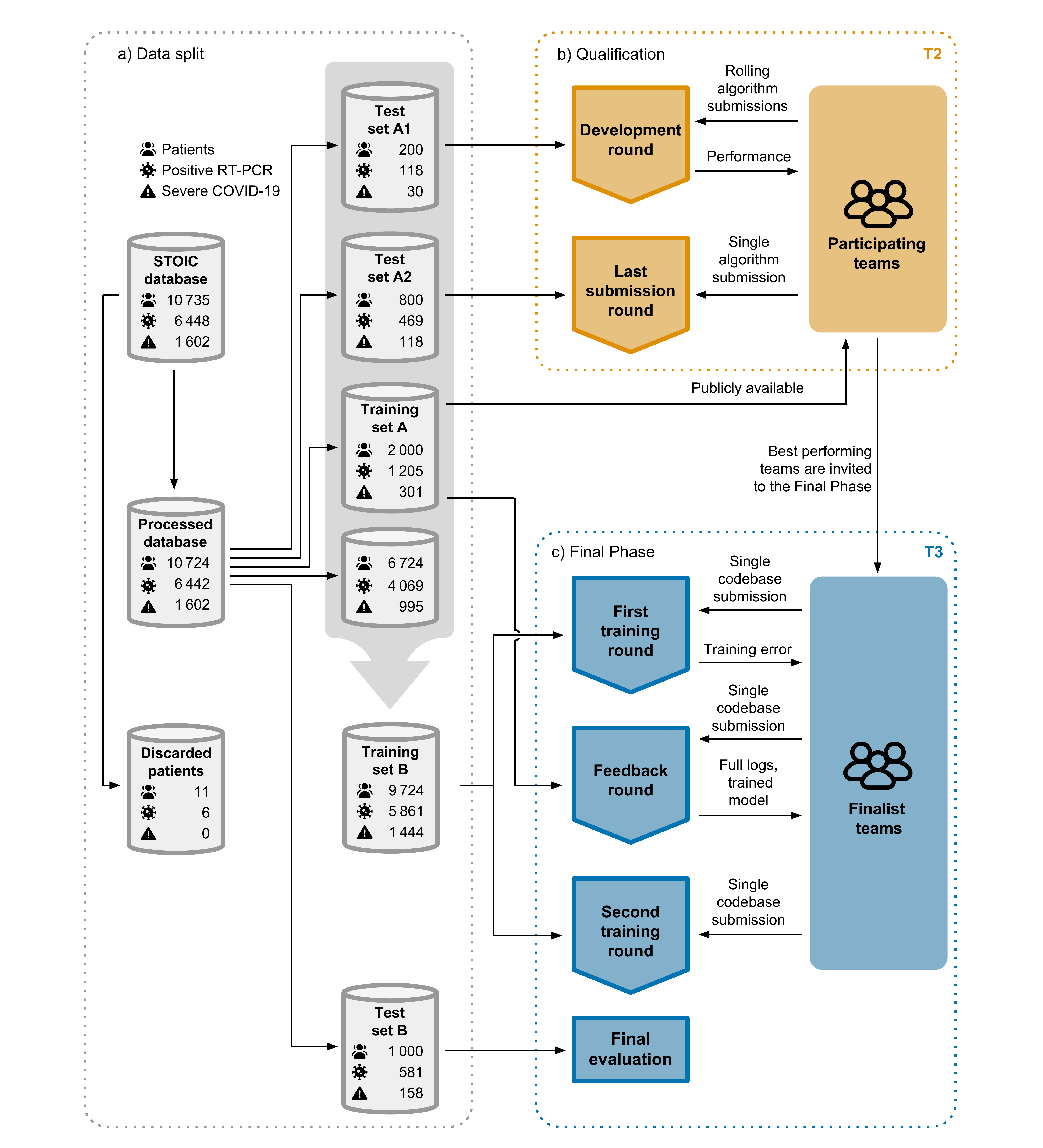}
  \caption{\small  Schematic overview of the STOIC2021 challenge. a) The CT scans in the STOIC database were discarded when severe motion artifacts that affected the entire scan were present, and preprocessed otherwise. From this processed database, training set A, and test sets A1, A2 and B were randomly sampled without replacement. Training set A, and test sets A1 and A2 were used in the Qualification phase. All processed data not present in test set B, including the 6\,724 CT scans not used in the Qualification phase, were used to form training set B. Training set B and test set B were used in the Final phase. All data except for the public training set A was kept secure on the grand-challenge.org platform at all times and could not be downloaded by participants at any point. The large sizes of test sets A2 and B were chosen to obtain accurate performance measures despite the class imbalance. Test set A1 was deliberately chosen to be smaller to lower the challenge organization costs of rolling submissions. b) In the T2 Qualification phase, participating teams trained challenge solutions on training set A and submitted them in a rolling fashion. They could view their performance on test set A1 through a public leaderboard. At the end of the Qualification phase a single submission for evaluation on test set A2 determined which teams were invited to join the Final phase. c) The T3 Final phase started with a first training round in which participants made a single codebase submission. The challenge organizers applied these codebases to training set B. The submitting teams received any training errors that their codebase generated. Subsequently, the finalist teams could make a Feedback codebase submission to resolve these errors. This codebase was applied to public training set A so that each finalist could inspect all results of their Feedback run. Lastly, finalists could submit their revised codebases to training set B, forfeiting their first training round submission. The models trained in the Final phase on training set B were evaluated on test set B.}
  \label{fig:2}
\end{figure*}

\section{Materials and methods}

\subsection{Materials}
Data from the STOIC study \cite{revel2021study} was used to construct the database used for the STOIC2021 challenge. For each subject in the database, the CT examination that was performed at presentation was selected. The subjects were represented by one thoracic CT scan when available, or one CT scan imaging more of the body otherwise. Slices more than 80 mm above and 110 mm below the lungs were discarded based on corrected lung masks produced by RTSU-Net \cite{Xie20}, as they were considered outside the typical scope of a thoracic CT scan.
For all subjects, sex and age labels, binned into ten year ranges, were provided as optional additional model input. RT-PCR results, and outcome, defined as death or intubation at one month, were used as ground truth for COVID-19 infection and severity respectively.  
Figure \ref{fig:2}a depicts how the preprocessed database was split into training and evaluation sets for the Qualification and Final phases of STOIC2021. 

\subsection{Performance metric}
Performance on all leaderboards was measured in terms of Area Under the receiver operating characteristics Curve (AUC) to reflect class imbalance \cite{reinke2021common}. Participants were ranked based on AUC for classifying COVID-19 severity, computed over cases with a positive COVID-19 RT-PCR result. AUC for COVID-19 presence, computed over all cases, was used solely as additional feedback for participants and did not directly influence ranking.

\subsection{Study design}
STOIC2021 was organized on the grand-challenge.org platform. It consisted of a Qualification phase followed by a Final phase as shown in Figure \ref{fig:2}. These phases respectively followed the T2 and T3 format illustrated in Figure \ref{fig:1}. Anyone with a verified, authentic user account on grand-challenge.org platform could join the challenge. Participants had the option to collaborate by forming non-overlapping teams. 

\subsubsection{Qualification phase}
During the Qualification phase, participating teams submitted solutions in the form of grand-challenge.org Algorithms trained on the publicly available training set A (see Figure \ref{fig:2}a), which was publicly released on December 6th, 2021.

\paragraph{Rolling submissions}
On December 23rd, a submission tutorial accompanied by a baseline system (1) was released and rolling submissions, which were evaluated on test set A1 (see Figure \ref{fig:2}a), were opened. This tutorial and source code is available on \url{https://github.com/luukboulogne/stoic2021-baseline}. Test set A1 consisted of only 200 subjects to limit the computational costs of the rolling submissions. Teams could view their performance on a public leaderboard. 
A count-down time between submissions of seven days was enforced. Violating this rule resulted in a submission time-out with a duration equal to the ignored count-down time. 

\paragraph{Last submission}
Teams submitted to test set A2 to qualify for the Final phase. To prevent the performance on the corresponding leaderboard to be tainted by overfitting, there existed no overlap between test set A1 and A2, and each team could submit their solution to be evaluated on test set A2 only once. 
Submissions to both test set A1 and A2 were closed on April 13th, 2022.

\subsubsection{Final phase}
The finalists were the 10 best performing teams that accepted an invitation to the Final Phase. They submitted code bases for performing training and inference with their solution.
A codebase for training and performing inference with the baseline system along with submission instructions for the Final phase was released on February 23rd, 2022. This tutorial and source code is available on \url{https://github.com/luukboulogne/stoic2021-baseline-finalphase}. These instructions ensured that the winning solutions could be used out-of-the-box by the challenge organizers and by third parties after the challenge had completed. 

The Final phase initially consisted of a single round in which the challenge organizers used the finalists' training code bases to train solutions. Since not all submissions completed training successfully during this first training round, the Final phase was extended with a feedback round and a second training round. 

Participating teams’ members qualified as author when submitting a codebase for training their solution to the Final Phase. Participating teams could publish their own results separately, without embargo.

\paragraph{Training environment}
\label{sec:trainingenvironment}
The training environment for the Final phase was drafted on March 17th based on resource requests and discussion with the Qualification phase participants, and was finalized on April 29th. Final phase training was performed on an Amazon EC2 p3dn.24xlarge instance. Each submission was allowed training for a maximum of 120 hours with access to two Tesla V100 GPUs with 32 GB vRAM each, 16 cpus with a total of 128G RAM, and 2 000 GB of Elastic Block Storage for storing intermediate results such as preprocessed data.

\paragraph{First training round}
Finalists could submit a single code base for training and inference with their solution in the form of a GitHub repository until May 12th. The challenge organizers generated training algorithms in the form of Docker \cite{merkel2014docker} container images from the submitted code bases and applied these to training set B (see Figure \ref{fig:2}). Each finalist obtained any error messages that their training algorithm generated in the first training round. These error messages were first scrutinized by the challenge organizers to ensure no leakage of sensitive information from training set B and to confirm the absence of indications of model performance. 

\paragraph{Feedback round}
To acquire additional feedback about running their code base in the training environment, finalists could submit any code base before July 17th following the final submission guidelines. These codebases were applied to the training environment and participants received the complete training logs and the resulting trained model. 
For the Feedback round only, two modifications were made to the training environment. Firstly, to ensure that training set B was kept secure, training set B was swapped out for the public training set A. Secondly, run time was limited to 24 hours to keep down computational costs.

\paragraph{Second training round}
Finalists were given the opportunity to make a second submission to the Final phase until July 27th. They could update their codebases to make their resulting training and inference containers run and complete successfully. For this update, methodological changes with respect to the first training round submission were not allowed. The codebases were checked for adherence to this rule by the challenge organizers and no violations were found. Finalists that chose to submit to the second training round were required to renounce their first training round submission. 

\subsubsection{Prizes}
Prizes in Amazon Web Services (AWS) credits were awarded to the best performing teams of the Final phase with values of \$10\,000, \$6\,000, and \$4\,000 for 1st, 2nd, and 3rd place respectively. The winners were announced during a public webinar on October 18th, 2022.

\subsubsection{Future submissions}
After STOIC2021 had concluded, rolling submissions to test set A1 were re-opened. The leaderboard corresponding to test set A2 was made available for submission upon request to the challenge organizers.

\subsection{Statistical tests}
The DeLong \cite{Delo88,sun2014fast} test is widely used for comparing AUCs and was also adopted for the statistical analysis in this work. 95\% confidence intervals were computed as the interval between the 2.5\% and 97.5\% percentiles of a bootstrap distribution generated with 1\,000 iterations \cite{moore1989introduction}.

\subsection{Baseline method}
The baseline for STOIC2021 implemented a simple training and evaluation pipeline for an Inflated 3D convnet (I3D) \cite{Carr17}.

\paragraph{Preprocessing strategy}
The input CT scans were resampled to an isotropic spacing of 1.6 mm3.  A center crop of 240*240*240 voxels was extracted from the CT, using zero padding when necessary. The voxel values were clipped between -1100 and 300 HU and rescaled to the range [0,1].

\paragraph{Training strategy}
A single I3D model \cite{Carr17}, initialized with publicly available weights trained for RGB video classification, was trained to estimate both COVID-19 presence and severity. The model was trained on all training data for 40 epochs using the AdamW optimizer \cite{LoshchilovHutter2018} with a learning rate of 10, momentum parameters $\beta_1$ = 0.9, $\beta_2$ = 0.99, and a weight decay of 0.01. Data augmentation was employed in the form of zoom, rotation, translation, and adding gaussian noise. Patient age and sex information were not incorporated as input to the model.

%% file: sections/results.tex
\begin{figure*}[htb]
  \centering
  \includegraphics[width=\textwidth]{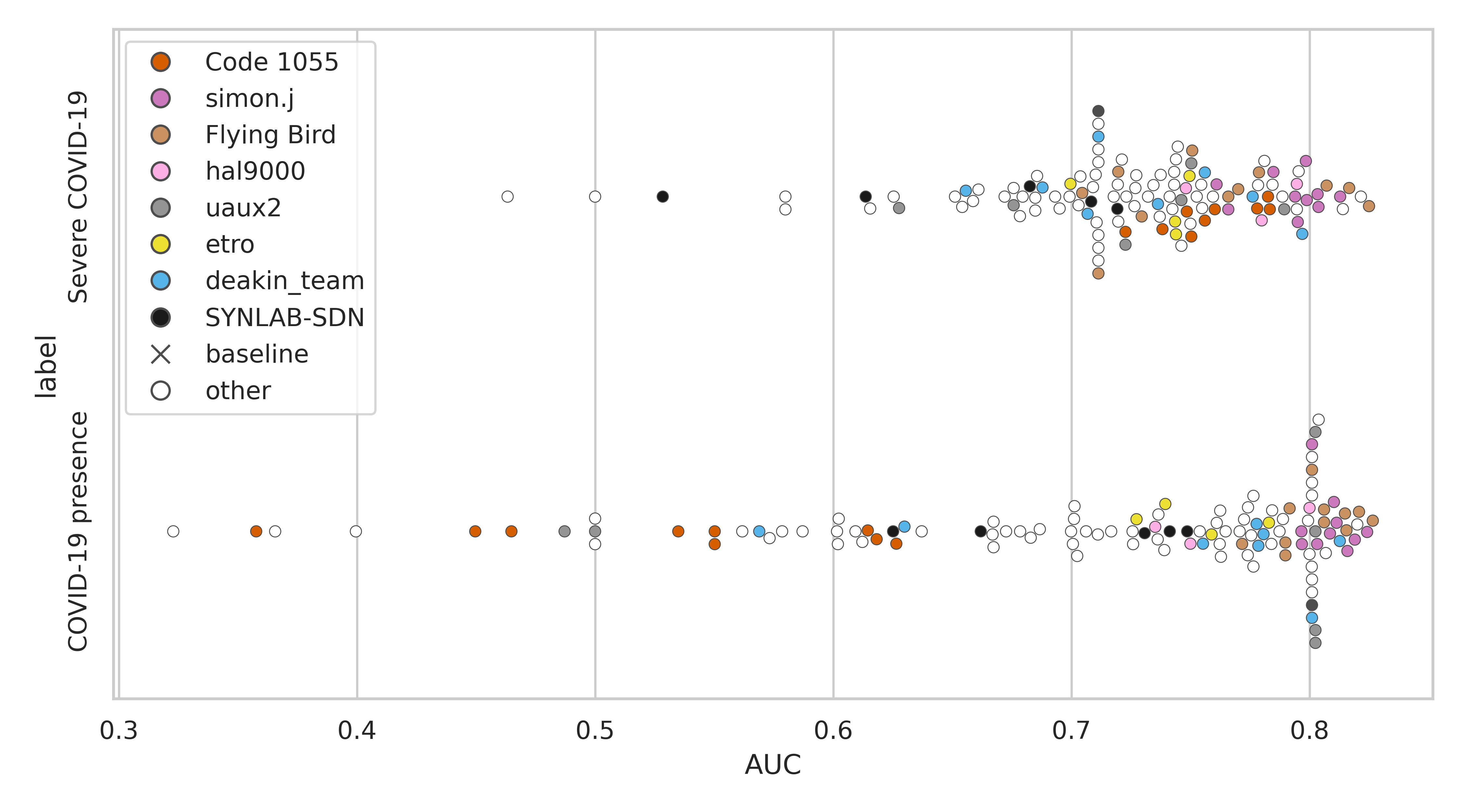}
  \caption{Performance distribution of the rolling submissions to test set A1 during the Qualification phase. The performance of the baseline is represented by an ‘x’. Submissions by the eight finalist teams are represented by colored circles. All other submissions are represented by white circles.}
  \label{fig:3}
\end{figure*}

\begin{table*}[!t]
    \caption{\label{tab:2}Performance on test set B. Solutions trained on training set A and B respectively are printed in regular and bold text. The top three ensemble was obtained by averaging the predictions of the best performing solutions, the AUCs of which are marked with ‘*’. }
    \centering
    \begin{tabular}{|r|l|l|}
        \hline
        Team name & AUC severe COVID-19 & AUC COVID-19 presence \\
        \hline
        \textbf{Top three ensemble} & \textbf{0.817} & \textbf{0.849} \\
        \textbf{Code 1055} & \textbf{0.815*} & \textbf{0.616} \\
        \textbf{simon.j} & \textbf{0.810*} & \textbf{0.845*} \\
        \textbf{Flying Bird} & \textbf{0.794*} & \textbf{0.838*} \\
        \textbf{hal9000} & \textbf{0.788} & \textbf{0.829*} \\
        \textbf{uaux2} & \textbf{0.787} & \textbf{0.825} \\
        \textbf{baseline} & \textbf{0.775} & \textbf{0.818} \\
        \textbf{etro} & \textbf{0.763} & \textbf{0.677} \\
        deakin\_team & 0.741 & 0.820 \\
        SYNLAB-SDN & 0.722 & 0.789 \\
        \hline
    \end{tabular}
\end{table*}
\input{sections/resultfigs}

\section{Results}

\subsection{Qualification phase}
413 participants registered to STOIC2021. During the rolling submissions, 30 teams, comprising 68 participants developed and successfully submitted 119 solutions to test set A1. Figure \ref{fig:3} shows an overview of the performance of these submissions. 20 teams competed for admission to the Final phase by successfully submitting to test set A2. The best performing teams on test set A2 were selected to advance to the Final phase, with invitations extended to the top ten teams that accepted.

\subsection{Final phase}
\subsubsection{First training round}
Eight of the ten Finalist teams submitted a codebase for training their solution on training set B. These eight teams are highlighted with unique colors in Figure \ref{fig:3}. In the first training round, the codebases submitted by the teams simon.j, Flying Bird, and etro completed successfully. All other codebases exited training with an error.

\subsubsection{Feedback round and second training round}
The teams Code1055, uaux2, and hal9000 submitted codebases to the feedback round and to the second training round. All three submissions to the second training round completed successfully, resulting in a total of six successful Final phase submissions. 

\subsubsection{Performance}
Table \ref{tab:2} shows the AUC on test set B for COVID-19 presence and severity of the teams that submitted to the Final phase. Figure \ref{fig:4} shows Receiving Operating Characteristics (ROC) curves of the six successful Final phase submissions for discriminating between severe and non-severe COVID-19 subjects from test set B.
Figures \ref{fig:5} and \ref{fig:6} show how the finalists ranked the subjects from test set B with severe and non-severe COVID-19 respectively for presence of severe COVID-19. Figures \ref{fig:7} and \ref{fig:8} highlight some individual cases from test set B. 
During the original STOIC project \cite{revel2021study}, a logistic regression model was developed to predict severe COVID-19 using clinical variables and CT annotations by radiologists. It was developed and evaluated using the patients from the STOIC who were COVID-19 positive for both RT-PCR and CT, and had unenhanced CT. Of these 4238 patients, 1000 developed severe COVID-19. Revel and colleagues 6 reported an AUC for this model of 0.69 (CI: 0.67-0.71). To compare this model against the results from STOIC2021, an ensemble of the top three solutions for severe COVID-19 prediction was evaluated on the 367 patients from test set B who were COVID-19 positive for both RT-PCR and CT, and had unenhanced CT. 97 of these patients developed severe COVID-19. The top three ensemble achieved an AUC of 0.783 (CI: 0.706-0.848).

\subsection{Solution methodology overview}
\label{sec:submittedmethods}
Most finalists used lung and/or lesion segmentation methods \cite{hofmanninger2020automatic,muller2021robust} to extract relevant features or to preprocess the input CT scan. Other preprocessing methods used were combinations of resampling, cropping, clipping, and normalizing or standardizing the image. End-to-end deep learning was the most common approach. The teams trained 2D or 3D versions of varying convolutional neural network architectures, \cite{liu2022convnet,He15b,howard2019searching,Huan17a} often starting from pre-trained weights, and using varying data augmentation methods. The finalists that did not employ end-to-end learning employed logistic regression on top of either processed features extracted by vision transformers \cite{dosovitskiy2020image} (simon.j) or features designed based on generated lung \cite{hofmanninger2020automatic} and lesion masks \cite{nvidia2023clara} (etro and SYNLAB-SDN). Compared to the end-to-end deep learning methods, these methods consumed less time and memory during training. Most teams used an ensemble of classifiers. The rest of this section contains a detailed overview of the methods that were successfully submitted to the Final Phase.

\subsubsection{Code 1055}
\input{sections/teammethods/code1055}

\subsubsection{simon.j}
\input{sections/teammethods/simonj}

\subsubsection{Flying Bird}
\input{sections/teammethods/flyingbird}

\subsubsection{hal9000}
\input{sections/teammethods/hal9000}

\subsubsection{uaux2}
\input{sections/teammethods/uaux2}

\subsubsection{etro}
\input{sections/teammethods/etro}

\subsubsection{deakin\_team}
\input{sections/teammethods/deakinteam}

\subsubsection{SYNLAB-SDN}
\input{sections/teammethods/synlabsdn}

%% file: sections/resultfigs.tex
\begin{figure*}[!p]
  \centering
  \includegraphics[width=.8\textwidth]{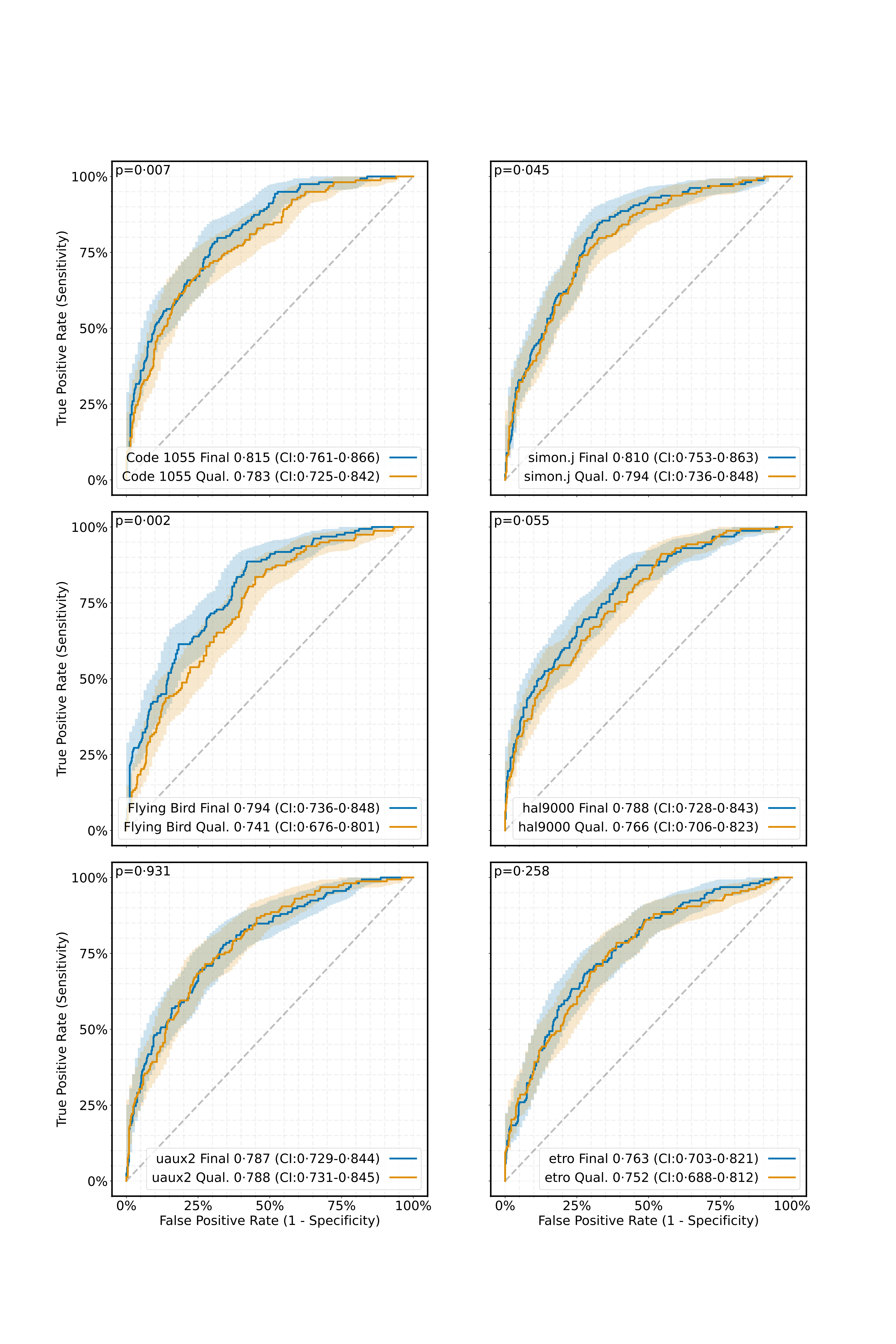}
  \caption{ROC curves with confidence intervals (CIs) for discriminating between severe and non-severe COVID-19 on test set B. The curves for the codebase submissions in the Final phase that completed training on training set B successfully are shown in blue. The ROC curves of the submissions that represented these teams in the Qualification phase, trained on training set A, are shown in orange. DeLong p-values are shown in the top left. AUCs with CIs are shown in the legends.}
  \label{fig:4}
\end{figure*}

\begin{figure*}[htb]
  \centering
  \includegraphics[width=\textwidth]{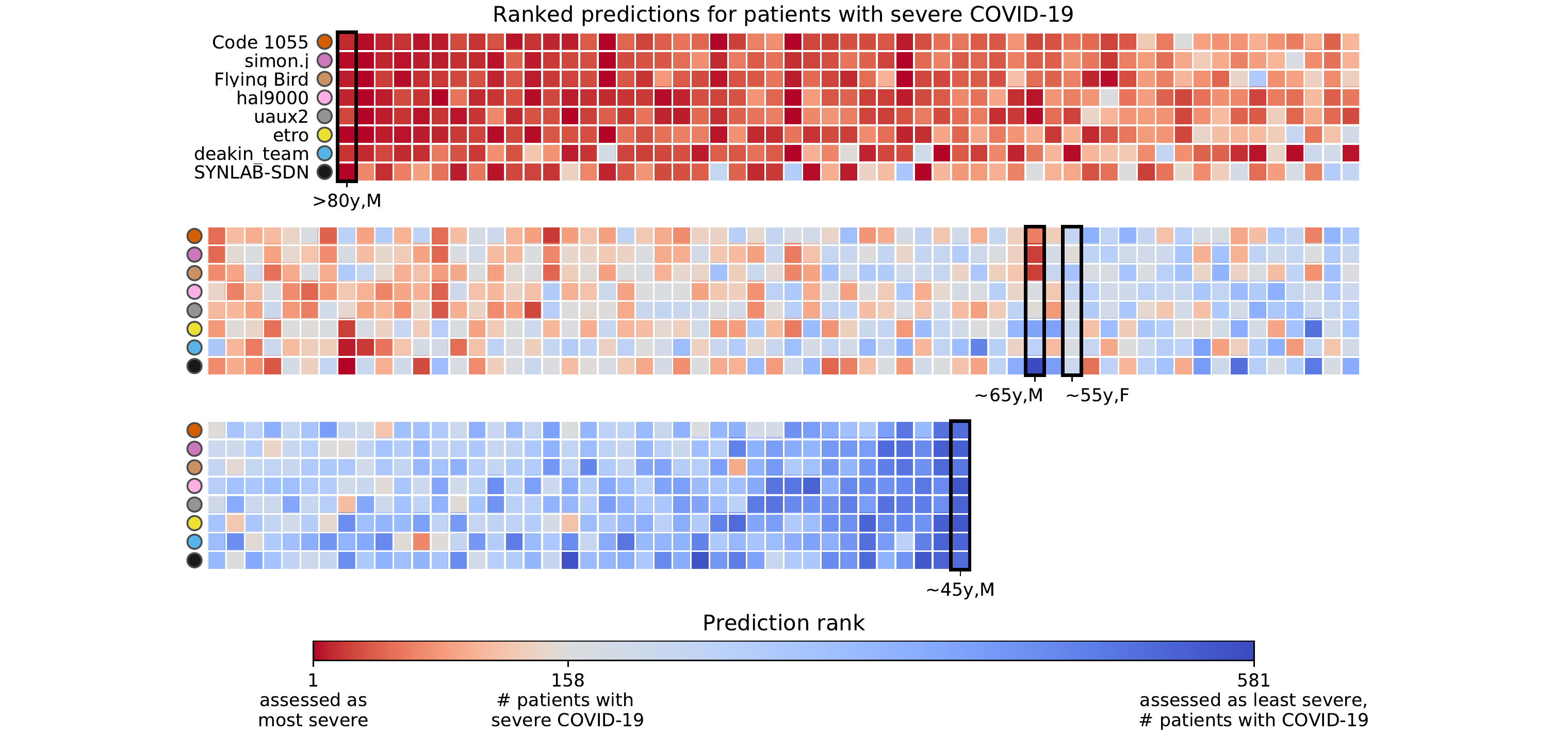}
  \caption{Ranked predictions for the subjects with severe COVID-19. Ranks were computed over all subjects from test set B with a positive RT-PCR test for COVID-19. Each column shows the ranked predictions of all finalist teams for one subject. The subjects are ordered by the average rank of all corresponding finalist predictions. Figure \ref{fig:7} shows the CT scans corresponding to the columns that are outlined in black and annotated with age and sex.}
  \label{fig:5}
\end{figure*}

\begin{figure*}[htbp]
  \centering
  \includegraphics[width=\textwidth]{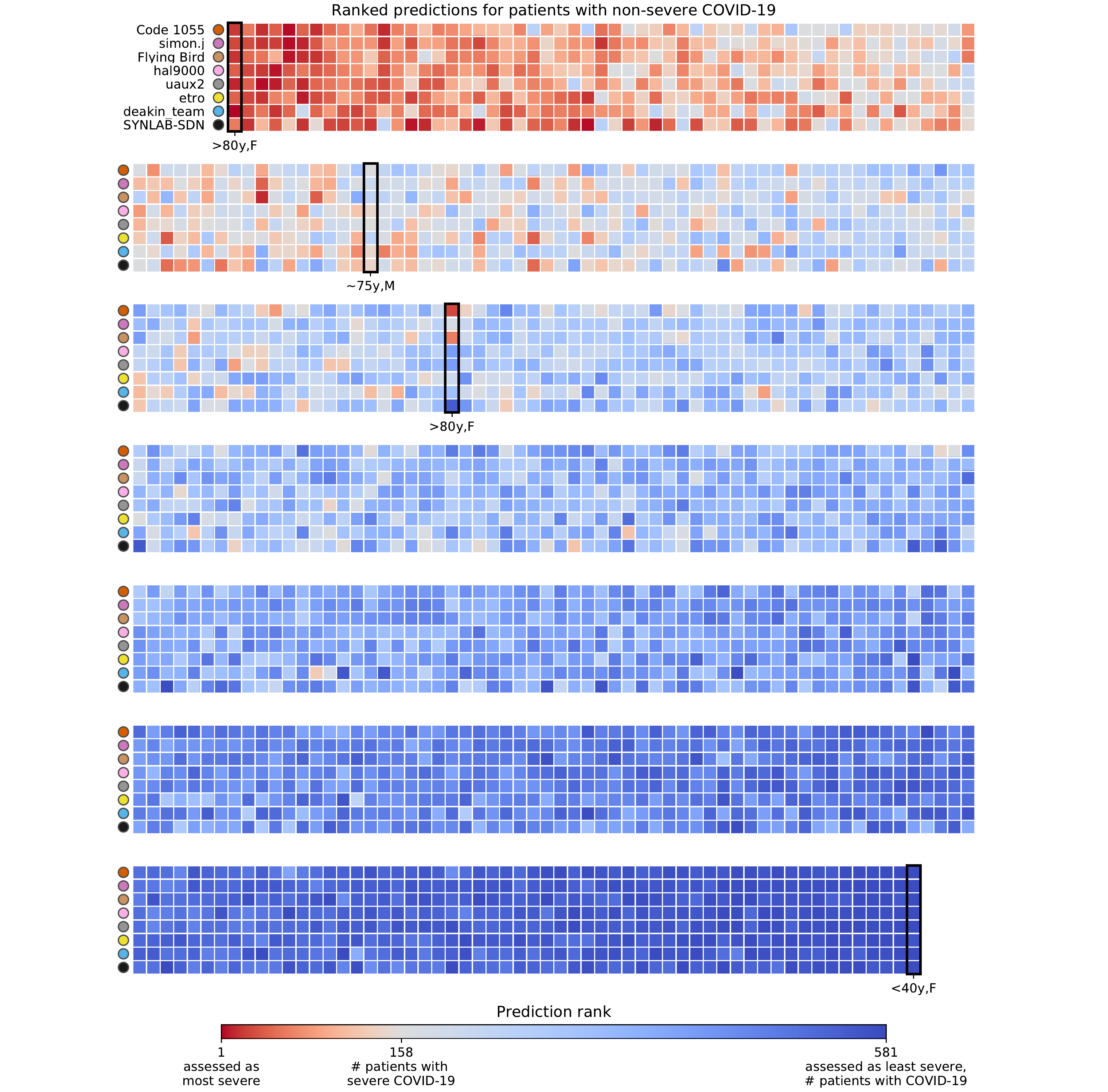}
  \caption{Ranked predictions for severe COVID-19 for subjects with non-severe COVID-19. Ranks were computed over all subjects from test set B with a positive RT-PCR test for COVID-19. Each column shows the ranked predictions of all finalist teams for one subject. The subjects are ordered by the average rank of all corresponding finalist predictions. Figure \ref{fig:8} shows the CT scans corresponding to the columns that are outlined in black and annotated with age and sex. }
  \label{fig:6}
\end{figure*}

\begin{figure*}[!p]
  \centering
  \begin{subfigure}{.8\textwidth}
    \centering
    \includegraphics[width=\textwidth]{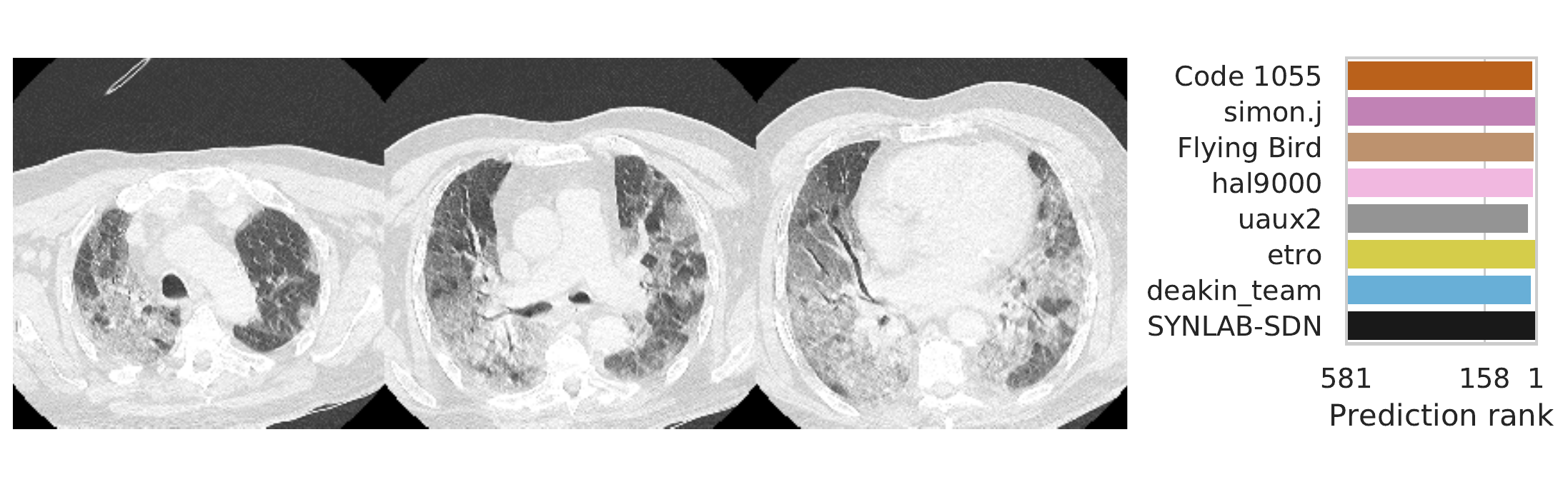}
    \caption{80+ years old, male}article
  \end{subfigure}
  \begin{subfigure}{.8\textwidth}
    \centering
    \includegraphics[width=\textwidth]{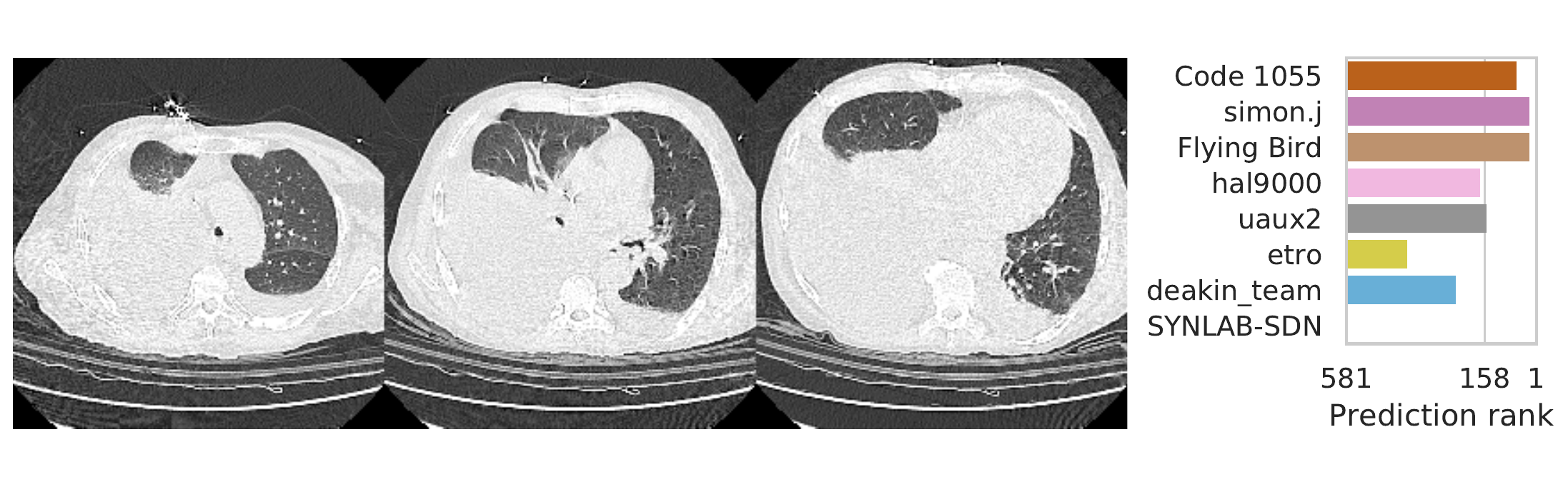}
    \caption{60-70 years old, male}
  \end{subfigure}
  \begin{subfigure}{.8\textwidth}
    \centering
    \includegraphics[width=\textwidth]{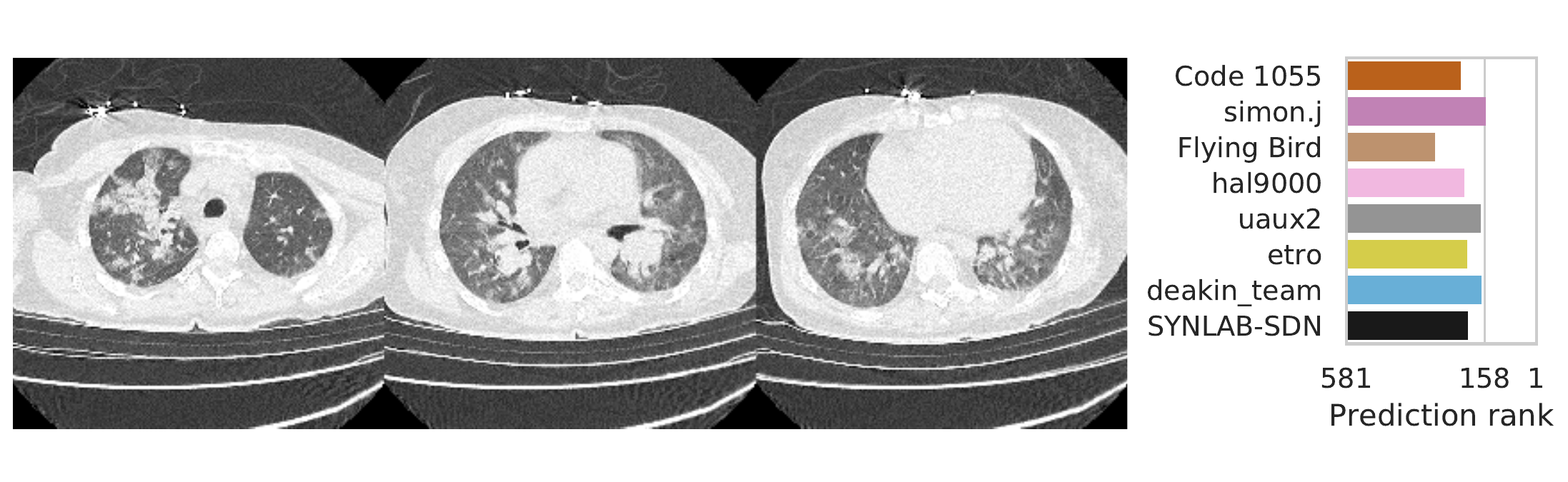}
    \caption{50-60 years old, female}
  \end{subfigure}
  \begin{subfigure}{.8\textwidth}
    \centering
    \includegraphics[width=\textwidth]{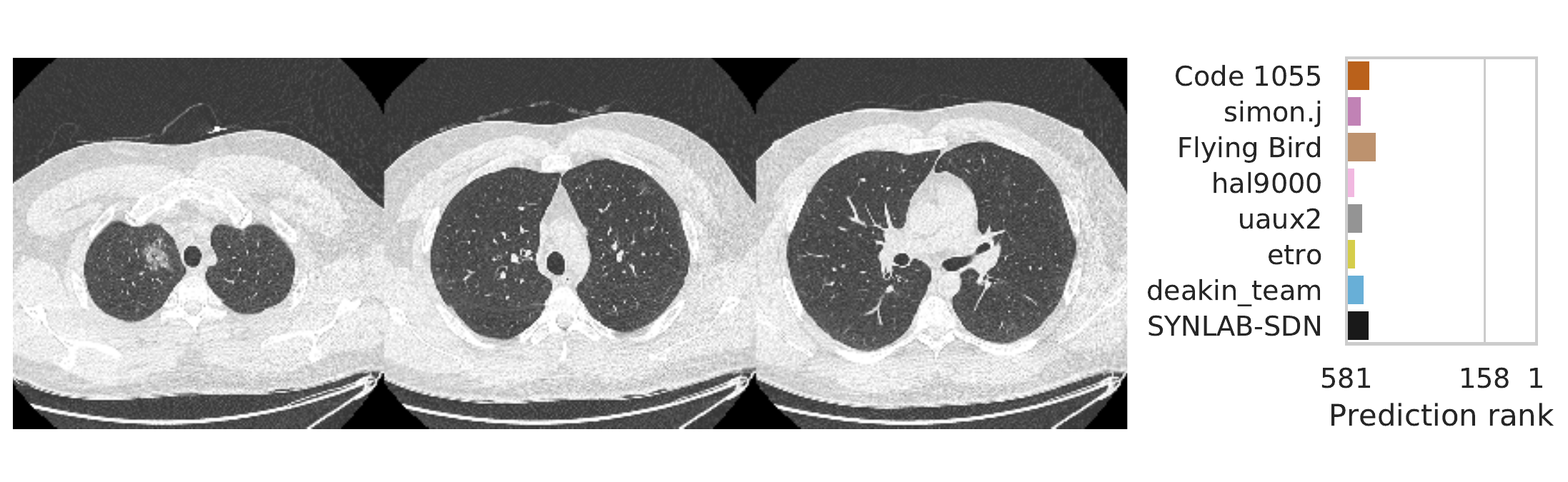}
    \caption{40-50 years old, male}
  \end{subfigure}
  \caption{Subjects from test set B with severe COVID-19 that were highlighted in Figure \ref{fig:5}. For each subject, three axial slices of a CT scan are shown on the left. The right shows how each finalist ranked the subject for presence of severe COVID-19. These ranks were computed over all subjects from test set B with a positive RT-PCR test for COVID-19.}
  \label{fig:7}
\end{figure*}

\begin{figure*}[!p]
  \centering
  \begin{subfigure}{.8\textwidth}
    \centering
    \includegraphics[width=\textwidth]{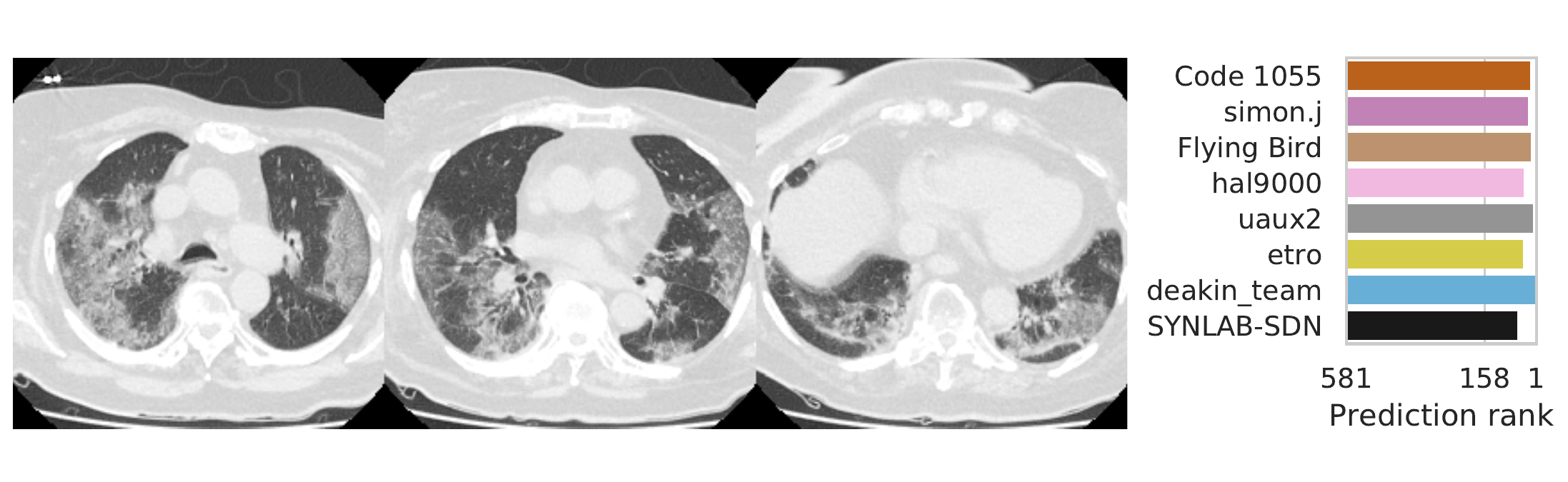}
    \caption{80+ years old, female}
  \end{subfigure}
  \begin{subfigure}{.8\textwidth}
    \centering
    \includegraphics[width=\textwidth]{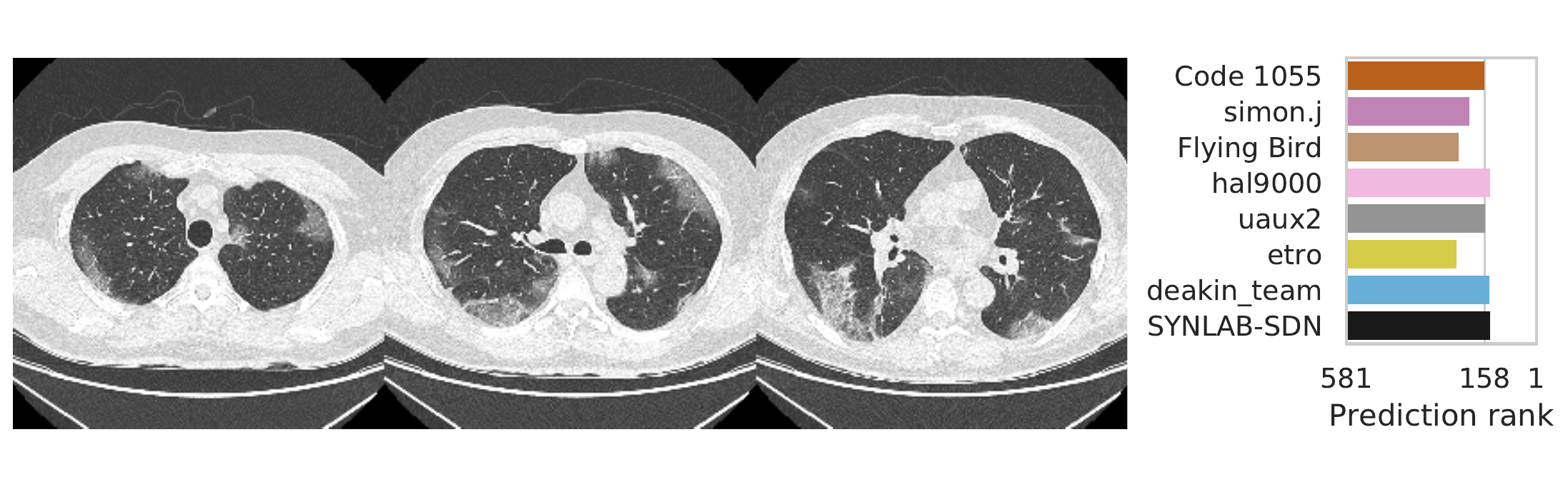}
    \caption{70-80 years old, male}
  \end{subfigure}
  \begin{subfigure}{.8\textwidth}
    \centering
    \includegraphics[width=\textwidth]{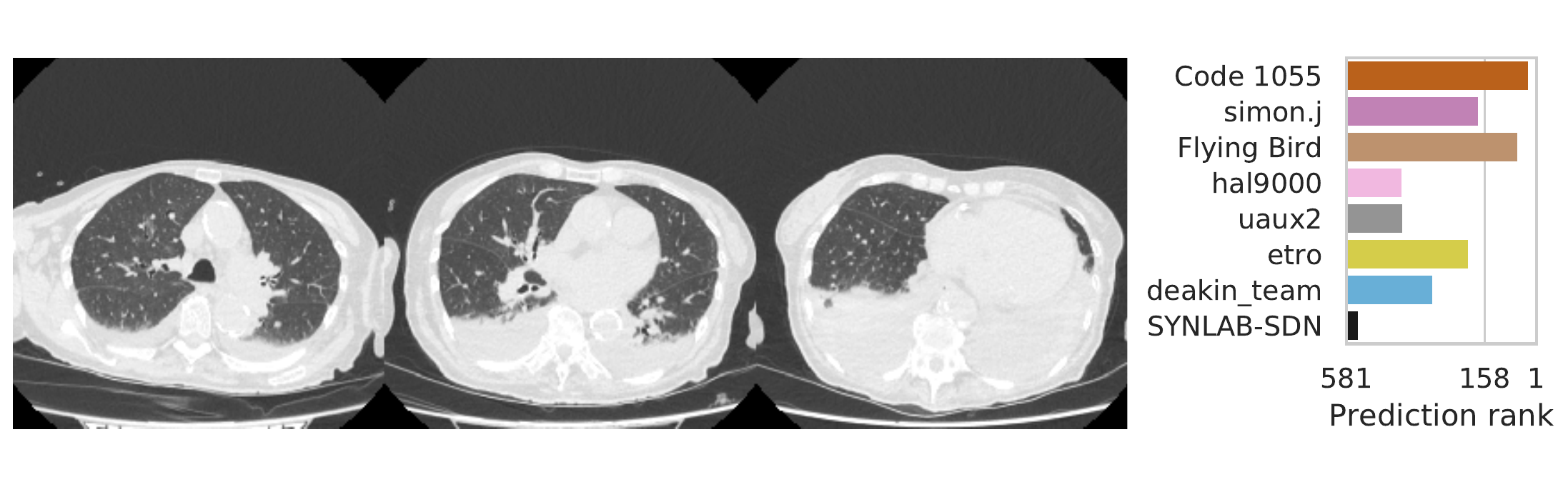}
    \caption{80+ years old, female}
  \end{subfigure}
  \begin{subfigure}{.8\textwidth}
    \centering
    \includegraphics[width=\textwidth]{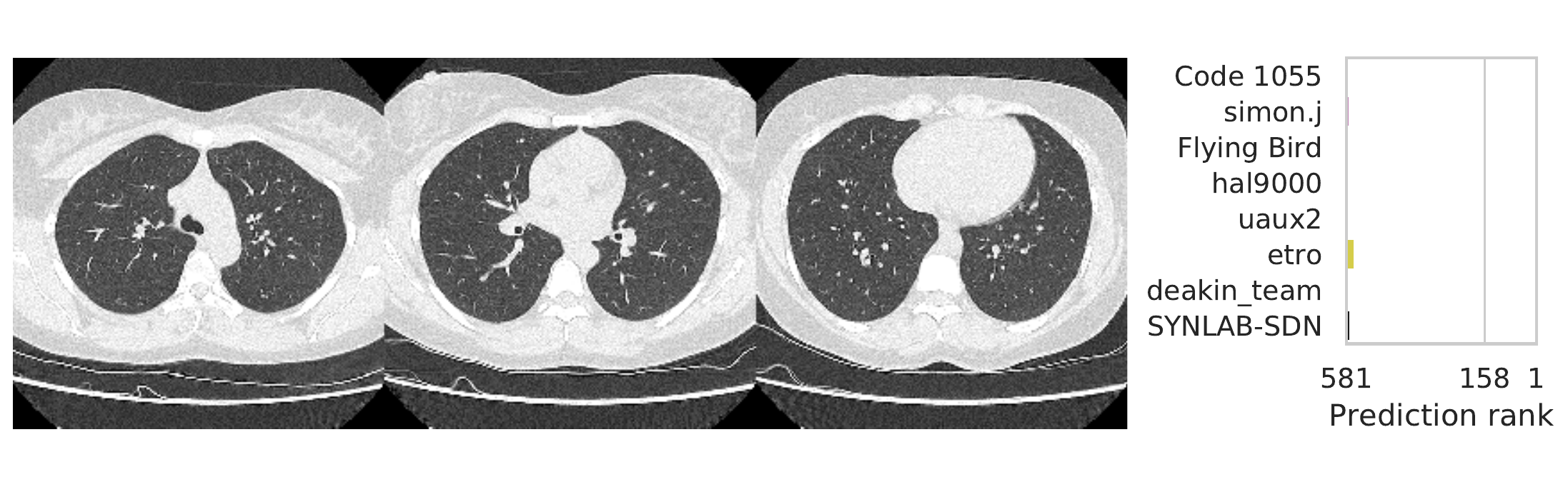}
    \caption{40- years old, female}
  \end{subfigure}
  \caption{Subjects from test set B with non-severe COVID-19 that were highlighted in Figure \ref{fig:6}. For each subject, three axial slices of a CT scan are shown on the left. The right shows how each finalist ranked the subject for presence of severe COVID-19. These ranks were computed over all subjects from test set B with a positive RT-PCR test for COVID-19.}
  \label{fig:8}
\end{figure*}

%% file: sections/teammethods/code1055.tex
\begin{figure*}[htb]
  \centering
  \includegraphics[width=.5\textwidth]{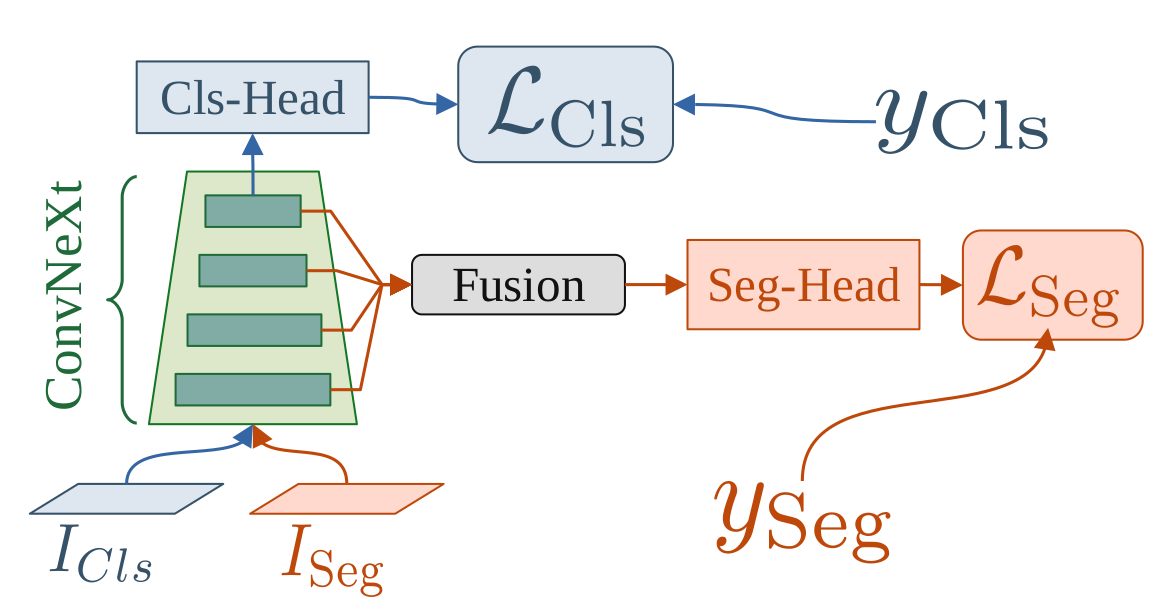}
  \caption{The pretraining pipeline is depicted. If segmentation data ($I_{Seg}$) is used as input, the features of each stage are upsampled, concatenated and the segmentation map is calculated with a segmentation head. If the classification data ($I_{Cls}$) is used as input, the severity prediction is obtained with a classification head using the features of the last stage. The overall loss is calculated as $L = L_{Cls} + L_{Seg}$}.
  \label{fig:code1055}
\end{figure*}

Severity classification using CT data is very similar to classical image classification apart from dealing with 3D tensors instead of 2D images. This allows us to employ the pre-existing techniques used in image classification. The ConvNeXt model \cite{liu2022convnet} combines the benefits of the modern Vision Transformers \cite{dosovitskiy2020image} with Convolutional Neural Networks (CNN) and thus reaches state-of-the-art ImageNet results. We implement – to the best of our knowledge – the first 3D version of this architecture and, thus, boost the performance for severity classification in contrast to conventional CNNs.

\paragraph{Preprocessing strategy}
The input CT scans were resized to $256\times256\times256 voxels$. Their intensity values were clipped between -1100 and 300 HU and normalized around zero with a standard deviation of one. 

\paragraph{Training strategy}
Even though the STOIC project \cite{revel2021study} is a comparably large database of CT scans, it is exceedingly small in contrast to ImageNet \cite{Russ14a}. Nevertheless, we are able to use a network with a large number of parameters and still prevent overfitting. For that purpose, we employ pretrained weights, a cosine learning rate scheduler, an early stopping strategy, an exponential moving average of the network parameters and efficient online data augmentation. Moreover, we balance our dataset in order to avoid learning a bias in the label distribution induced by the small number of severe cases.

In order to initialize our model with useful weights, we pretrain our network on two additional datasets. First, we train a 2D ConvNeXt on grayscale images from ImageNet. We calculate a superposition of gaussian inflated 2D weights to obtain 3D ImageNet weights. To further adjust these inflated ImageNet weights to our three dimensional task, we perform an additional multitask-pretraining using a segmentation \cite{roth2022rapid,an2020ct,Clar13} and classification \cite{morozov2020mosmeddata} dataset. We use an architecture inspired by UPerNet \cite{xiao2018unified} to concurrently perform segmentation of the infected lung region for the segmentation data and prediction of severity for the classification data. This pre-training scheme is depicted in Figure \ref{fig:code1055}. We are able to increase the performance of our model significantly with this additional pretraining in contrast to randomly initialized weights or inflated ImageNet weights.

In order to prevent overfitting and achieve greater generalization we use online data augmentation to virtually increase the dataset size. Besides using standard transforms like flipping, rotation or cropping, we apply a novel implementation of elastic deformations. By separating the gaussian kernels and utilizing GPU hardware, we are able to perform extremely fast elastic deformations. Consequently, we can augment our data with almost no additional cost. Furthermore, we perform 5-fold cross-validation during training.

Follow-up work is published by \cite{kienzle2023covid}.

\paragraph{Inference strategy}
We average the outputs of the 5 networks trained in the cross validation. Therefore, we are able to train with the complete dataset and still generalize very well.

\paragraph{Public access}
Public code for training and inference publicly available at \url{https://github.com/DIAGNijmegen/stoic2021-finalphase-submission-code1055}. Algorithm available for public use at \url{https://grand-challenge.org/algorithms/code-1055-second-final-phase-submission/}.

%% file: sections/teammethods/simonj.tex
Balaitous is an updated version of the AI-severity algorithm \cite{lassau2021integrating} implemented in the scancovia repository \cite{Jegou2022}. Given an input CT scan, the model outputs a probability for COVID-19 disease and for severe outcome (intubation or death within one month).

\paragraph{Preprocessing strategy}
The CT scan was rescaled to a resolution of 1.5 mm $\times$ 1.5 mm $\times$ 5 mm and reshaped to a
shape of 224 $\times$ 224 $\times$ $D$. A lung segmentation mask was comput
\paragraph{Summary}ed using a 2D U-Net \cite{hofmanninger2020automatic} and cleaned. The scan was cropped to the slices containing the lungs. 
For each slice, a first feature vector $X_{full}$ was extracted using a ViT-L model. This model was pretrained on ImageNet-22k using iBOT \cite{zhou2021ibot} and fine-tuned for 35 epochs on 165k CT scan images from 4k patients and 7 datasets. Next, the lung mask was applied so that only the lungs were visible and a second feature vector Xlung was extracted using the same ViT-L model without fine-tuning. For both VIT-L models, the extracted features of the individual slices were combined through pixel-wise average pooling.

\paragraph{Training strategy}
For the severe outcome two logistic regressions were applied to [$X_{full}$, age, sex] and [$X_{lung}$, age, sex]. The two predictions were aggregated through a learned weighted average. For the COVID-19 presence two logistic regressions were applied to $X_{full}$ and $X_{lung}$ and the two predictions were aggregated through a learned weighted average. 
Training was performed in 32 folds in the form of four different eight-fold cross validations.

\paragraph{Inference strategy}
The predictions were combined linearly with weights optimized that maximize the performance on the 32 training folds.

\paragraph{Methods altered from Qualification phase to Final phase}
None.

\paragraph{Public access}
Public code for training and inference publicly available at \url{https://github.com/SimJeg/balaitous} and \url{https://github.com/DIAGNijmegen/stoic2021-finalphase-submission-simonj}. Algorithm available for public use at \url{https://grand-challenge.org/algorithms/simonj-first-final-phase-submission/}.

%% file: sections/teammethods/flyingbird.tex
The method employed was end-to-end deep learning with ResNet18 \cite{He15b} models.

\paragraph{Preprocessing strategy}
In order to minimize image size and eliminate irrelevant regions, an open source lung segmentation model \cite{hofmanninger2020automatic} was employed. The lung masks were used to crop the images, and were expanded by 6 mm to ensure complete coverage. The resulting cropped images were rescaled to 256 $\times$ 256 $\times$ 256 voxels using trilinear interpolation. The voxel values were then clipped to the range (-1024, 512), and standardized with a mean of -237 and a standard deviation of 404.

\paragraph{Training strategy}
Due to the substantial volume of data, training a 3D network from scratch without a pre-trained model would be time-consuming. Regrettably, there is no all-purpose pre-trained model suitable for 3D networks. As a result, our approach involves initially training a pre-trained model via self-supervision \cite{zhou2019models}, followed by conducting classification tasks built upon the pre-trained model.
We used cross validation. For training each fold, we appended a decoder to the ResNet18 network. Then, following the method described in \cite{He15b}, we applied some transformations to the input image and fed the transformed image into the network. We trained the network to enable it to recover the original image from the transformed image. After training, we obtained a pre-trained ResNet18 model. In the subsequent COVID-19 classification task and severity task, we initialized our models using pre-trained ResNet18.
For both the COVID-19 classification task and severity task, we employed the same data augmentation techniques, including rotation, scaling, flipping, elastic transformation, Gaussian noise, and Gaussian smoothing. We used cross-entropy loss function and AdamW \cite{LoshchilovHutter2018} optimizer, along with a one-cycle learning rate policy.
For the severity task, we also incorporated age information by concatenating the age, which was divided by 100, with the output of ResNet18, thereby taking into account the influence of age on severity. Furthermore, the data used in this task only consisted of COVID-19 positive cases.

\paragraph{Inference strategy}
For each model obtained through the cross-validation, test time augmentations are applied. The original input image is passed through the model, as well variants of it obtained by flipping along each of the three axes, obtaining four outputs per model. Finally, the outputs of all models are averaged to obtain the final output.

\paragraph{Methods altered from Qualification phase to Final phase}
The data augmentation methods underwent minor modifications. The severity model was trained using both COVID-19 negative and positive images during the qualification phase, whereas only COVID-19 positive images were utilized in the final phase. Combinatorial image flipping was applied for test time augmentation during the qualification phase, along each of the three axes, resulting in a total of 8 outputs per model (2 $\times$ 2 $\times$ 2). In the final phase, only 4 outputs were generated, including the original image and those flipped along the x, y, and z axes.

\paragraph{Public access}
Public code for training and inference publicly available at \url{https://github.com/DIAGNijmegen/stoic2021-finalphase-submission-flyingbird}. Algorithm available for public use at \url{https://grand-challenge.org/algorithms/flying-bird-first-final-phase-submission/}.

%% file: sections/teammethods/hal9000.tex
We employed an ensemble of ResNet18 \cite{He15b}, and MoblieNetV3-Large \cite{howard2019searching} models trained end to end to predict COVID-19 disease and severity. In each model, embeddings of all 32 slices were averaged and passed through a classifier to get the disease and severity probabilities. The ensemble of multiple models was used by averaging the probabilities of each model. 

\paragraph{Preprocessing strategy}
32 equidistant slices were sampled from the input CT scan. These slices were resampled to 224 $\times$ 224 pixels. The pixel values were clipped between -1350 and 150 HU. The images were normalized to  a mean of 0.5 and a standard deviation of 0.5.

\paragraph{Training strategy}
The data for model development was split ten times into a training and validation set, such that the training set contained 85\% of the data. A ResNet18 \cite{He15b} was trained on five of these splits, and a MobileNetV3-Large \cite{howard2019searching} was trained on the other five. Before presenting input data to a model, data augmentations were applied in the form of resizing, horizontal flipping, random cropping, gamma correction, color jitter, rotation, and blurring. The embeddings of all 32 slices were averaged and passed through a classifier to get the disease and severity probabilities. All models were trained using the Adam optimizer, with a learning rate of 0.0001 and weight regularization of 0.0005. The learning rate decayed by a factor of 0.1 every 40 epochs.

\paragraph{Inference strategy}
All model predictions were combined through averaging. We employed extensive test time augmentations involving five different crops (four corner crops and the center crop), and three different rotations (minus five degrees, plus five and plus ten degrees), and averaged the predictions for each augmentation. This was done for all five models for each model class. The ensemble prediction was obtained by averaging the probabilities.

\paragraph{Methods altered from Qualification phase to Final phase}
In the Qualification phase, we trained an ensemble of only MobileNet V3 Large models.

\paragraph{Public access}
Public code for training and inference publicly available at \url{https://github.com/DIAGNijmegen/stoic2021-finalphase-submission-hal9000}. Algorithm available for public use at \url{https://grand-challenge.org/algorithms/hal9000-second-final-phase-submission/}.

%% file: sections/teammethods/uaux2.tex
\begin{figure*}[htb]
  \centering
  \includegraphics[width=.7\textwidth]{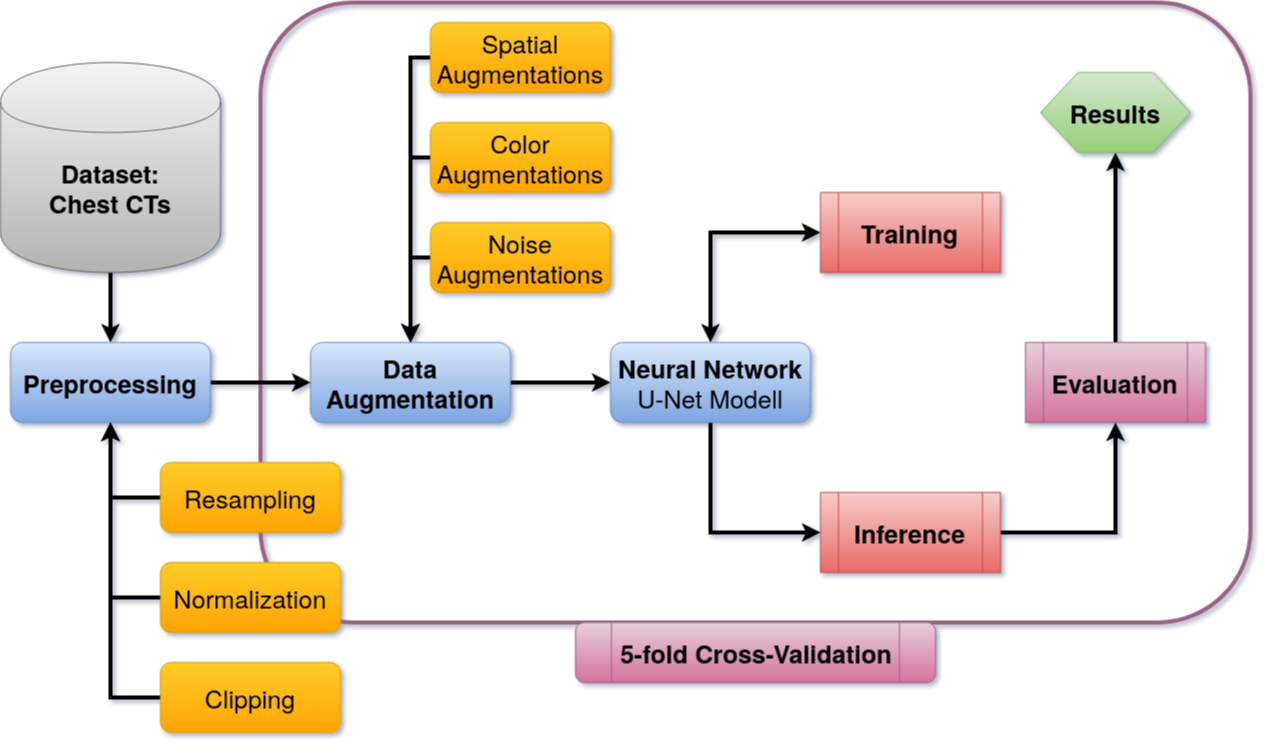}
  \caption{The MIScnn pipeline for SARS-CoV-2 segmentation to calculate the Infection-Lung-Ratio \cite{muller2021robust}}
  \label{fig:uaux2}
\end{figure*}

To assess the severity of SARS-CoV-2 (COVID-19) based on Computed Tomography (CT) scans of the lung, we apply an ensemble method approach, where we combine meta-data and 3D-CNN predictions. In addition to the information on patient age and sex already present in the data set, we rely on the respective Infection-Lung-Ratio (ILR) to generate our predictions. For implementation, we used our in-house developed framework AUCMEDI which is built on TensorFlow \cite{MullerKramer}.

\paragraph{Preprocessing strategy}
For preprocessing, first, all data samples were resampled to a voxel spacing of 1.48 $\times$ 1.48 $\times$ 2.10 and clipped to the range [-1024, 100] to exclude irrelevant Hounsfield Unit areas \cite{yamada2022visual}. Subsequently, the data was standardized to grayscale. Training samples that might exceed the accepted input image size of 148 $\times$ 224 $\times$ 224 were either randomly cropped or zero-padded to match the required size. For inference, center cropping was applied. To enable transfer learning, the grayscale images were converted to RGB. The intensities were scaled to the range of [0, 1]. Then, normalization was applied via the Z-Score normalization approach based on the mean and standard deviation computed on the ImageNet dataset \cite{Deng09}.

\paragraph{Training strategy}
In line with current state-of-the-art approaches, we applied several augmentation methods on the dataset, including rotation, flipping, scaling, gamma modification, and elastic deformations. 
Our main model for COVID-19 Severity prediction is based on a custom 3D version of the DenseNet121 architecture. We modified the classification head to additionally take metadata into account, which is described later on. For the training process, we applied transfer learning on the classification head and a fine-tuning strategy on all layers. The transfer learning is done for 10 epochs, using the Adam optimizer with an initial learning rate of $1 \times 10^{-4}$ and a batch size of 4. The fine-tuning runs for a maximum of 240 epochs, using a  dynamic learning rate starting from $1 \times 10^{-5}$ to a maximum decrease to $1 \times 10^{-7}$ (decreasing factor of 0.1 after 8 epochs without improvement on the monitored validation loss). Furthermore, an early stopping technique was utilized, stopping after 36 epochs without improvement. As a loss function, we utilized the weighted Focal loss \cite{Lin17a}. For inference, the model with the best validation loss is used. For COVID-19 presence prediction, we utilize a model based on the 3D ResNet34 architecture with the same hyperparameter settings as described above, that predicts 3 classes (negative/positive/severe). COVID-19 presence equals the sum of positive and severe cases. The metadata consists of three parts: Patient age, sex, and the ILR of each sample. The latter describes the ratio between infected parts of the lung and healthy tissue. We calculate the ILR by feeding the data into the MIScnn segmentation framework \cite{muller2021robust,muller2021miscnn}, which utilizes a standard U-Net to predict infected areas Figure \ref{fig:uaux2}. For COVID-19 severity prediction, we applied cross-validation with a dynamic number of folds as a bagging approach for ensemble learning and monitored the outputs on the validation loss. We aimed to create a variety of models which were trained on different subsets of the training data. 

\paragraph{Inference strategy}
Our final COVID-19 severity prediction comprises the averaged sum of all five predictions. This approach not only allows for a more efficient usage of the available training data but also increases the reliability of the prediction.

\paragraph{Methods altered from Qualification phase to Final phase}
In the Qualification phase, cross-validation was done with five folds.

\paragraph{Public access}
Public code for training and inference publicly available at \url{https://github.com/DIAGNijmegen/stoic2021-finalphase-submission-uaux2}. Algorithm available for public use at \url{https://grand-challenge.org/algorithms/uaux2-second-final-phase-submission/}.

%% file: sections/teammethods/etro.tex
A short-term COVID-19 severity classifier was developed through logistic regression considering age, sex, and several image-derived features. A previously trained lung lesion segmentation model was used to extract volume fractions for ground glass opacities and consolidations. The segmentations were used in combination with the CT scan to derive mean intensities, kurtosis, and skewness for healthy lung parenchyma and lesion tissue. The final severity prediction was made by an ensemble of 20 models, trained on covid-positive samples selected through bootstrapping with replacement.

\paragraph{Preprocessing strategy}

The lungs were segmented using an open-source segmentation model \cite{hofmanninger2020automatic}.  A postprocessing step was added retaining only the 2 largest components and setting a minimum size for the components to exclude any regions outside the lungs that may have been segmented. CT scans were cropped to the lung mask and resampled to an isotropic spacing of 1 mm. The intensities were clipped to [-1000 HU, 100 HU] and scaled to [-1, 1]. Ground glass opacity and consolidation patterns were segmented using a previously trained lung lesion segmentation model. The nnU-Net implementation in Monai \cite{Monai} was used. The hyperparameters for this deep learning pipeline were determined automatically using the heuristics developed in nnU-Net \cite{Isen19}. The network was trained using the sum of the mean dice loss and the cross entropy, and deep supervision. Training data included 199 CT scans of the COVID-19 lesion segmentation challenge \cite{CovidSegmentation}, 69 scans and manual lung lesion segmentations from the icovid consortium \cite{ICovidAI}, 70 scans from the COPLENet public dataset \cite{COPLENet} and 10 scans from the publicly available COVID-19 CT Lung and Infection Segmentation Dataset \cite{Ma2020}. From these lung and lesion segmentations, the lesion volume fractions were calculated by dividing the lesion volume by the total lung volume. Additionally, the mean intensity, kurtosis and skewness were derived for each type of lesion and the healthy lung tissue.

\paragraph{Training strategy}
A logistic regression was trained for severity. Patient age and sex categories were assigned numerical values and were complemented with several image-derived features. Volume fractions of ground glass opacity and consolidation were included, as well as the mean intensity, kurtosis and skewness for healthy lung parenchyma and both lesion classes separately. For patients that were considered lesion free, the intensities and textural features of the ground glass opacity and consolidation were given the values of the healthy tissue. All intensity  features were rescaled to [-1, 1].
To improve robustness, the severity classifier was built up by bagging 20 models where each training set was composed using bootstrapping with replacement on the covid-positive samples.

\paragraph{Inference strategy}
For inference, the intensity features were rescaled using the corresponding extrema from the training set. Final probabilities for severe COVID-19 were obtained by averaging the predictions of the 20 models. The probability of COVID-19 was predicted by a previously trained 3D ConvNext \cite{liu2022convnet} model.

\paragraph{Methods altered from Qualification phase to Final phase}
For the Qualification phase, the model for severity was trained on both COVID-19 positive and negative patients versus only positives for the Final phase. For COVID-19 presence detection, the ConvNext model was added in the Final phase while a regression model similar to the severity classifier was used for the Qualification.

\paragraph{Public access}
Public code for training and inference publicly available at \url{https://github.com/DIAGNijmegen/stoic2021-finalphase-submission-etro}. Algorithm available for public use at \url{https://grand-challenge.org/algorithms/etro-first-final-phase-submission/}.

%% file: sections/teammethods/deakinteam.tex
The method employed was end-to-end deep learning with DenseNet-201 \cite{Huan17a}.
\paragraph{Preprocessing strategy}
The input CT scans were resampled to an isotropic spacing of 1.6 mm3.  A center crop of 240 $\times$ 240$\times$ 240 voxels was extracted from the CT, using zero padding when necessary. The voxel values were clipped between -1100 and 300 HU and rescaled to the range [0,1].

\paragraph{Training strategy}
A 3D DenseNet-201 \cite{Huan17a}, initialized with weights trained on the public STOIC2021 training set, was trained using the Adam optimizer with a learning rate of 0.00004, and a batch size of two for 15 epochs.

\paragraph{Inference strategy}
Inference was performed by a forward pass through the trained DenseNet-201 model.

\paragraph{Methods altered from Qualification phase to Final phase}
An ensemble approach incorporating multiple models, specifically DenseNet-201, DenseNet-169, and DenseNet-121, was initially proposed for this study. However, due to constraints related to computational resources and time in the training environment, we were ultimately only able to train a DenseNet-201 model.

\paragraph{Public access}
Algorithm available for public use at \url{https://grand-challenge.org/algorithms/baseline-13/} (Qualification phase submission).

%% file: sections/teammethods/synlabsdn.tex
The method was based on logistic regression using patient age, sex and features extracted from lesion masks.

\paragraph{Preprocessing strategy}
The CT voxel intensity values were clipped to the range [-1000, 500]. Afterward, a pre-trained model for COVID-19 lesion segmentation by Nvidia Clara (2) was used to obtain suitable masks representative of COVID-19 lesion burden. Furthermore, a lung mask was segmented from the input CT scan using a U-Net \cite{hofmanninger2020automatic}.
From the lesion masks, the following features were extracted:
\begin{itemize}
    \item Mean HU value,
    \item Standard deviation intensity,
    \item Percent of lesion volume, computed as lesion volume divided by lung volume,
    \item Number of connected components in the lesion mask.
\end{itemize}

In addition, patient age and sex were included as features. 
0
\paragraph{Training strategy}
For the classification, the dataset was randomly split into a training/validation (80\%) and testing set (20\%). Z-normalization was applied to the features constituting the training set, and the mean and standard deviation values calculated on the training set were used on the validation and test set. A downsampling strategy was applied to balance the dataset. We have trained logistic regression to solve the tasks. K-Fold cross-validation with K = 5 was applied to the training dataset for model selection in the form of hyperparameter tuning.

\paragraph{Inference strategy}
The trained logistic regression model was applied to perform inference.

\paragraph{Methods altered from Qualification phase to Final phase}
None.

\paragraph{Public access}
Algorithm available for public use at \url{https://grand-challenge.org/algorithms/2steps-2/} (Qualification phase submission).

%% file: sections/discussion.tex
\section{Discussion}
The Type Three (T3) medical image analysis challenge format presented in this study allows solutions to be trained on private data and that guarantees that their training methodologies are reusable. T3 was implemented in the STOIC2021 challenge, in which participants predicted from an initial CT scan, whether a COVID-19 patient would be intubated or would die within one month. 

To evaluate their solutions, challenges typically release test set images to enable participants to run inference on them \cite{ouyang2019analysis,Anto21,Ehte17,lassau2020three,choi2022challenge,halabi2019rsna,ali2021deep,knoll2020advancing,porwal2020idrid,kim2021paip,fang2022adam,sun2021multi,sathianathen2022automatic,combalia2022validation,kavur2021chaos,hakim2021predicting,Hell19a,Bogu19,orlando2020refuge,Yang18a,hirvasniemi2023knee,arganda2015crowdsourcing,ivantsits2022detection,caicedo2019nucleus,simoes2020bciaut,Veta18,winzeck2018isles,marinescu2019tadpole,balagurunathan2021lung,de2021generalizability,bratholm2021community,bron2015standardized,Seti17,pan2019improving,cash2015assessing,kim2020challenge,qsm2021qsm,fu2020age,babier2021openkbp}. STOIC2021 consisted of a Qualification phase that instead followed the structure implemented of some recent challenges \cite{Bult22,Aubr22,schirmer2021neuropsychiatric,da2022digestpath,sun2022crowdsourcing,hatt2018first} where participants submit solutions trained on public data, and of a T3 Final phase. The Final phase solutions consistently outperformed the solutions submitted to the Qualification phase by the same participants, indicating that T3 may improve challenge solution performance through training on a combination of public and private data.

STOIC2021 resulted in six publicly available codebases through which the training and inference methods for the top performing solutions can be accessed. The challenge organizers tested these codebases by training the corresponding solutions without manual intervention by the participating teams. This guaranteed the reusability by third parties of these publicly released training methodologies. Links to these codebases can be found in section \ref{sec:submittedmethods}.

Advanced age and male sex are risk factors for severe outcome of a COVID-19 infection \cite{revel2021study}.
Most finalists used sex and age information as additional input to their model. Besides employing a pre-trained segmentation model, most of the submitted solutions use 3D image processing methods that are not specific to one task or image modality. The released training methods may be useful for any 3D medical image classification tasks.

STOIC2021 participants were not incentivized to focus on the confirmation of COVID-19 presence, since this is possible with high sensitivity through RT-PCR testing \cite{tsang2021diagnostic}. The absence of this incentive explains why team Code 1055, which achieved the highest AUC for discriminating between severe and non-severe COVID-19 in the Final phase, achieved the lowest AUC for detecting COVID-19 presence of all finalists. It also explains why, overall, the finalists’ performances on the auxiliary metric of detecting COVID-19 presence did not align with the finalists’ ranks in the Final phase.

This study has limitations. Participants of STOIC2021 were not incentivized to focus on the calibration or interpretability of their solutions. Also, datasets for externally validating solutions on their ability of predicting intubation or death within one month were not publicly available. This also prohibited directly comparing the presented performances to the algorithms trained to predict severe COVID-19 outcome by \cite{lassau2021integrating}. However, the solution by simon.j was heavily based on this work. Furthermore, T3 challenges are limited by the computational budget of the challenge organizers. STOIC2021 therefore implemented a limit to the compute resources for training the Final phase solutions, as detailed in section \ref{sec:trainingenvironment}, and allowed for a limited number of finalists. Lastly, the maximum obtainable performance is limited by imperfections in the COVID-19 severity and presence labels. Death at one month follow-up could have resulted from any cause. RT-PCR is an imperfect ground truth for infection. For the STOIC study, 39\% of initially negative RT-PCR tests were found to be positive when repeated in patients with typical clinical signs of COVID-19 \cite{revel2021study}.

\section*{Conclusion}
This work showed the efficacy of the T3 medical image analysis challenge format. T3 improves upon the formats of previous challenges in two main ways. Firstly, it allows challenge solutions to be trained on private data. This results in training on bigger data, which can increase the performance of the resulting challenge solutions. Secondly, it ensures that the training methods developed for the challenge can be used out-of-the box by third parties.